\documentstyle[preprint,pre,aps,eqsecnum,12pt,epsf]{revtex}

\newcommand{\ewxy}[2]{\setlength{\epsfxsize}{#2}\epsfbox[-40 60 640
590]{#1}}
\newcommand{\real}{{\sf I}\kern-.12em{\sf R}}

\begin{document}

\draft
\preprint{IFUP-TH 26/96}
\draft
\title{Topological Susceptibility at zero and finite $T$ 
in $SU(3)$ Yang-Mills theory\footnote {Partially 
supported by EC Contract CHEX-CT92-0051 and by MURST.}}
\author{B. All\'es, M. D'Elia and A. Di Giacomo}
\address{Dipartimento di Fisica dell'Universit\`a and INFN, 
Piazza Torricelli 2, 56126-Pisa, Italy}
\maketitle
\begin{abstract}
We determine the topological susceptibility $\chi$ at $T=0$ in pure 
$SU(3)$
gauge theory and its behaviour at finite $T$ across the deconfining
transition. We use an improved topological charge density operator.
$\chi$ drops sharply by one order of magnitude at the deconfining
temperature $T_c$.
\end{abstract}

\vskip 5mm


\vfill\eject

\section{Introduction}

The flavour singlet axial current
\begin{equation}
j^{\mu}_5 =  \bar{\psi} \gamma^5 \gamma^{\mu} \psi
\end{equation}
is not conserved in QCD because of the triangle anomaly \cite{hoft1}
\begin{equation}
\partial_{\mu} j^{\mu}_5 (x) = 2 N_f Q(x). 
\label{eq:anomaly}
\end{equation}
In eq. (\ref{eq:anomaly})  $Q(x)$ is the topological 
charge density, defined as
\begin{equation}
Q(x) = {{g^2} \over {64 \pi^2}} \epsilon^{\mu \nu \rho \sigma} 
 F^a_{\mu \nu} (x) F^a_{\rho \sigma} (x).
\end{equation}
As 
a consequence the corresponding $U_A (1)$ is not a symmetry \cite{hoft1}.

The non singlet partners
\begin{equation}
j^{\mu}_{5 a} =  \bar\psi \gamma^5 \gamma^{\mu} \lambda_a \psi
\end{equation}
are conserved, and the corresponding symmetry is spontaneously broken, 
the pseudoscalar octet being the Goldstone particles. 

If $U_A (1)$ were a symmetry, either parity doublets should exist, or, in case
of spontaneous breaking, the inequality $m_{\eta '} \le \sqrt{3} m_{\pi}$ 
should hold \cite{wein}.
Neither of these predictions is true in nature, and this has been known as 
the $U_A(1)$ problem for many years, before the advent of QCD. 

However $U_A(1)$ is a symmetry at leading order in the expansion in 
$1 \over {N_c}$ \cite{wit1}, ($N_c$ is the number of colours). There are 
arguments that 
the leading approximation in that expansion describes the main physics
 of QCD \cite{hoft2,vene1}.
 The anomaly is non leading, and can be viewed as
a perturbation. One of its effects is to displace $m_{\eta '}$ from zero, 
which corresponds to the Goldstone particle in the leading order 
approximation, by an amount which is related 
to the topological susceptibility $\chi$ of the vacuum at the leading 
order. The prediction is \cite{wit2,vene2}
\begin{equation}
  {{2 N_f} \over {f_{\pi}^2}} \chi = m_{\eta }^2 + m_{\eta '}^2 
                       - 2 m_{K}^2.
\label{eq:massform}
\end{equation}
The topological susceptibility $\chi$ is defined as 
\begin{equation}
 \chi \equiv \int d^4 x \langle 0 | T(Q(x) Q(0))| 0 \rangle. 
\end{equation}
Leading order implies 
absence of fermions and in the language of the lattice
this is known as quenched approximation. Lattice is the ideal tool to 
compute $\chi$ from first principles. 
$\chi$ has in fact been determined \cite{digia1} 
and is consistent with the prediction 
of ref. \cite{wit2}. An additional hint in favour of it is the 
indication 
that the $\eta '$ mass is higher in sectors with higher topological charge 
\cite{fuku}. 

A question then arises naturally whether
 the $U_A (1)$ symmetry is restored
in quenched QCD at the same temperature at which $SU_A (3)$ is 
restored, i.e. at $T_c \sim 260$ MeV \cite{boyd}. 
Many models \cite{shu1} predict the behaviour of the $U_A(1)$ 
chiral symmetry at 
$T_c$. A quite general expectation is that the topological susceptibility
should drop at $T_c$ \cite{pisar}, since
Debye screening inhibits tunneling between states of different 
chirality and damps the density of instantons. 

Attempts have been made a few years ago to study the behaviour of $\chi$ 
through $T_c$ \cite{tep2,digia2}. The status is discussed in ref. \cite{digia2}. 
The difficulties go back to the definition of a topological charge 
on the lattice. The correct way to define it, according to the commonly 
accepted prescriptions of field theory, is to introduce on the lattice
a local operator $Q_L(x)$ for the topological charge density 
which tends to the continuum 
operator as the lattice spacing goes to zero. $Q_L$ provides a regularized
 version of $Q(x)$. In going to the continuum limit a proper renormalization  
must be performed, like in any other regularization scheme. A specific 
feature of $Q_L$ is that on the lattice it is not the divergence of a current,
 like in the continuum, and hence it renormalizes multiplicatively: this means
that the lattice topological charge of a configuration can be non integer 
\cite{z1}. In formulae  
\begin{equation}
Q_L = Z(\beta ) Q a^4 + {\cal O}( a^6 ).
\end{equation}
As usual, $\beta \equiv 6/g_0^2$.
The topological susceptibility can be defined on the lattice as 
\begin{equation}
\chi_L \equiv \langle \sum_x Q_L(x) Q_L(0) \rangle.  
\end{equation}
The standard rules of renormalization then give
\begin{equation}
 \chi_L = Z(\beta)^2 a^4 \chi + M(\beta) + {\cal O}(a^6),
\label{eq:defchilat}
\end{equation}
where  $M(\beta)$ is an additive renormalization containing
mixings of $\chi_L$ to other operators with the same  quantum numbers
and lower or equal dimensions \cite{tutti}. In formulae
\begin{equation}
M(\beta) = B(\beta) a^4 G_2 + P(\beta).
\end{equation}
The terms proportional to
$P(\beta)$ and $B(\beta)$ are respectively the mixings to the identity 
operator and to the density of action $G_2 \equiv
\langle {g^2 \over {4 \pi^2}} F_{\mu\nu}^a F_{\mu\nu}^a \rangle$. The 
additive renormalization comes from the singularities of the product 
$Q(x) Q(0)$ as $x \rightarrow 0$ and must be removed to be consistent 
with the prescription used to derive eq. (\ref{eq:massform}) \cite{crew}. 
The definition of $Q_L$ is not unique: infinitely many operators can
be defined which obey eq. (1.7) but differ by terms of order 
${\cal O}(a^6)$.
The simplest definition of $Q_L$ is \cite{divec}
\begin{equation}
Q_L(x) = {{-1} \over {2^9 \pi^2}} 
\sum_{\mu\nu\rho\sigma = \pm 1}^{\pm 4} 
{\tilde{\epsilon}}_{\mu\nu\rho\sigma} \hbox{Tr} \left( 
\Pi_{\mu\nu}(x) \Pi_{\rho\sigma}(x) \right).
\end{equation}
Here ${\tilde{\epsilon}}_{\mu\nu\rho\sigma}$ is the
standard Levi-Civita tensor for positive directions while for negative
ones the relation ${\tilde{\epsilon}}_{\mu\nu\rho\sigma} =
- {\tilde{\epsilon}}_{-\mu\nu\rho\sigma}$ holds. 
$\Pi_{\mu\nu}$ is the plaquette in the $\mu - \nu$ plane.
With this definition
$Z \simeq 0.18$ and the mixing 
$M$ is large compared to the signal in the scaling region \cite{digia3}. 
$Z$ and $M$ can be computed 
non-perturbatively \cite{np1,np3bis,np3}. 
Although the field-theoretic method is correct in principle, 
it is unpleasant that most of the signal is due to lattice artifacts, 
which have then to be removed. Moreover at 
the time of ref. \cite{digia2} the non-perturbative 
determination of $Z$ and $M$ was not known. 

An alternative method to determine $\chi$ is the so called cooling 
technique \cite{tep2}: the idea is to freeze 
quantum fluctuations by a local 
algorithm which cools the links one after the other. The modes relevant at 
a distance $d$ are frozen after a number of steps $n$, which is proportional
to $d^2$, like in a diffusion process. Most of the instantons 
are expected to have a size of the order of the correlation length. After 
a few cooling steps, the elimination of local fluctuations 
will suppress
the mixing $M$ and make  $Z \simeq 1$, but the 
number of instantons will be 
preserved, so that $\chi_L \simeq \chi a^4 $. 
In fact a plateau is reached in $Q_L$ after a few cooling steps, 
where $Q_L$ is an integer, which 
stands many further steps \cite{tep2}. At $T = 0$ and below 
$T_c$ the method works very well and agrees with the field theoretical 
method \cite{tutti}. Approaching $T_c$ from below, however, the plateau 
becomes shorter and shorter, collapsing to a maximum, 
which at higher $T$ decreases to non integer values.
No unambiguous criterion can then be given to determine the value of $\chi$.
The origin of the above behaviour is well understood.
A finite temperature $T$  on a lattice is obtained by taking a size 
$N_s^3 \times N_{\tau}$ 
($N_s$ spatial size, $N_{\tau}$ time size), with $N_{\tau} \ll N_s $.
The temperature is given by 
\begin{equation}
 T =  {{1} \over {N_{\tau} a(\beta)}}.
\end{equation}
As $T$ rises to $T_c$ the 
correlation length becomes equal to  $N_{\tau} a$, and the instantons, being 
of the size  of the correlation length, become  infrared unstable.

A third method to determine $\chi$ is the so called geometrical method
\cite{lus,kron}
by which a lattice configuration is interpolated by a continuous 
configuration on which the topological charge is read. In this way the charge
is  always an integer, which hopefully should coincide with the true 
topological charge in the continuum limit when $\beta \rightarrow \infty$ and 
the correlation length goes large. This hope, however, is not 
realized: at  $\beta \rightarrow \infty$ lattice artefacts dominate 
\cite{tep3}, at least for the usual actions. No determination of $\chi$ at 
finite $T$ 
 exists in the literature by this method. On the lattice actions can be 
constructed \cite{hperfect}
which belong to the same class of universality of Wilson action, 
for which dislocations could be absent: a proof, however, is still lacking, 
that the topological charge determined by the geometrical method 
coincides with the continuum value as defined in textbook field theory. 
Moreover it is not clear how the susceptibility computed with the
geometrical charge compares with the prescription of ref. \cite{crew}
for the singularity at $x=0$. Even if dislocations were absent a
mixing with the gluon condensate, which scales with the same power 
up to corrections due
to the dependence of $B(\beta)$ on $\beta$,
could still be present and should somehow be removed. 
Historically the method was proposed because eq. (\ref{eq:defchilat}) 
had been used to 
estimate $\chi$, but the factor $Z^2$ had been neglected in the analysis 
\cite{divec}, thus leading to a value of $\chi$ much smaller than what required 
by eq. (\ref{eq:massform}). 

Recently \cite{christou} an improved operator for $Q_L$ has been defined, 
which 
has the correct continuum limit (1.7), but is nearer to the 
continuum, in the sense that $Z$ is near to 1, and the mixing 
of $\chi_L$ to the 
identity is negligible in the scaling region. The operator has been tested
successfully in $SU(2)$ gauge theory with the normal Wilson 
action. 

What we do in the present paper is to implement the field theoretical
method using the improved operator for $Q_L$. Also for $SU(3)$ we find
that lattice artifacts are drastically reduced with respect to the
original choice, eq. (1.11), and become unimportant. They are anyhow
removed non-perturbatively.

We redetermine $\chi$ at $T = 0$, 
and study its behaviour through $T_c$. The new determination of $\chi$ at 
$T = 0$ is consistent with previous determinations \cite{digia3,np3,tepchi}.
At finite $T$ our main result is that $\chi$ drops at $T_c$ 
by more than one order of magnitude. 
In sect. II the results are presented and discussed. 
The details of the determinations of $Z(\beta)$ and 
$M(\beta)$ are described in sect. III.
Sect. IV contains a few concluding remarks.

\section{The method}
We will compute the lattice topological susceptibility $\chi_L$ 
defined in equation (\ref{eq:defchilat}) by use of the improved operators
defined in ref. \cite{christou} for the topological charge density
\begin{equation}
Q_L^{(i)}(x) = {{-1} \over {2^9 \pi^2}} 
\sum_{\mu\nu\rho\sigma = \pm 1}^{\pm 4} 
{\tilde{\epsilon}}_{\mu\nu\rho\sigma} \hbox{Tr} \left( 
\Pi^{(i)}_{\mu\nu}(x) \Pi^{(i)}_{\rho\sigma}(x) \right).
\end{equation}
In this definition, 
$\Pi^{(i)}_{\mu\nu}$ is the plaquette constructed with  $i$ times smeared
links $U^{(i)}_{\mu}(x)$. They are defined as
\begin{eqnarray}
U^{(0)}_{\mu}(x) &=& U_{\mu}(x), \nonumber \\
{\overline U}^{(i)}_{\mu}(x) &=& (1-c) U^{(i-1)}_{\mu}(x) +
{c \over 6} 
\sum_{{\scriptstyle \alpha = \pm 1} \atop { \scriptstyle 
|\alpha| \not= \mu}}^{\pm 4}
U^{(i-1)}_{\alpha}(x) U^{(i-1)}_{\mu}(x+\hat{\alpha})
U^{(i-1)}_{\alpha}(x+\hat{\mu})^{\dag}, \nonumber \\
U^{(i)}_{\mu}(x) &=& {{{\overline U}^{(i)}_{\mu}(x)} \over
{ \left( {1 \over 3} \hbox{Tr} {\overline U}^{(i)}_{\mu}(x)^{\dag} 
{\overline U}^{(i)}_{\mu}(x) \right)^{1/2} } }.
\end{eqnarray}

The improving is a local smearing inspired by the usual cooling procedure 
\cite{tep2}.
The parameter $c$ can be tuned to optimize the improvement. 
The formal continuum limit of $Q_L^{(i)}$ 
is $Q_L^{(i)} {\buildrel {a \rightarrow 0} \over \longrightarrow} 
a^4 Q + {\cal O}(a^6)$ for any $i$.
In our simulations we shall make use of the 0,1 and 2-improved
topological charge density operators $Q_L^{(0)}(x)$, $Q_L^{(1)}(x)$ and
$Q_L^{(2)}(x)$. 
Correspondingly we will compute the topological susceptibility
\begin{eqnarray}
\chi_L^{(i)} &\equiv& \langle \sum_x Q_L^{(i)}(x) Q_L^{(i)}(0) \rangle
\nonumber \\
 &=&  Z^{(i)}(\beta)^2 a^4 \chi + M^{(i)}(\beta) + {\cal O}(a^6), 
\label{eq:defrenchii}
\end{eqnarray}
where
\begin{equation}
M^{(i)}(\beta) \equiv B^{(i)}(\beta) a^4 G_2 + P^{(i)}(\beta).
\end{equation}

In eq. (\ref{eq:defrenchii}), $\chi$ and $G_2$ do not depend 
on the operator used for $Q_L$;
$Z^{(i)}(\beta)$, $B^{(i)}(\beta)$ and $P^{(i)}(\beta)$ do.
We checked that $Z^{(i)}$ does not depend within errors on the lattice
size, as expected from the fact that its value is determined
predominantly by short range fluctuations at the UV cutoff.
The multiplicative and additive renormalizations will be determined by
a non-perturbative technique \cite{np1,np3bis,np3} as follows.
To determine $Z^{(i)}$ we perform a few updating 
steps on a classical configuration consisting of one instanton, and we measure
$Q^{(i)}_L$ at each step: as long as no new instantons or antiinstantons are 
produced and the initial instanton is present, the topological charge of the 
configuration is preserved and is equal to the  known charge $Q$ of the 
instanton. 
A plateau will be reached when the local fluctuations which produce 
renormalization will thermalize, and on the plateau 
$Q^{(i)}_L = Z^{(i)} Q$, whence $Z^{(i)}$ can be determined. 

To determine $M^{(i)}(\beta)$ again we start from a configuration of known 
$Q$, e.g. the flat configuration (with all links $U_{\mu}(x)={\sf 1}
$) where $Q = 0$. 
We then start producing from it a sample of configurations by the usual 
updating procedure used to thermalize. We measure at each step 
$\chi^{(i)}_L$ and as 
long as no instanton or antiinstanton is created or destroyed, after a few 
steps local fluctuations will be thermalized and a plateau will be reached, 
according to eq. (\ref{eq:defrenchii}): the first term 
$Z^{(i)}(\beta)^2 a(\beta)^4 Q^2$ 
is known (it is zero if the initial configuration is flat) and
$M^{(i)}$ can be extracted.
A careful check shows that $M^{(i)}(\beta)$ determined in this way is 
independent of the $Q$ of the initial configuration: $Q = 0$ and $Q=1$ 
configurations give the same value of $M^{(i)}(\beta)$, within errors. 
In the continuum formulation of the theory this corresponds to the 
well-known fact 
that short distance effects, like renormalizations, do not depend 
appreciably on the classical background. 

To be sure that the number of instantons is not changed by the heating 
procedure, each configuration of the sample is checked during heating by  
performing a few cooling steps to detect its topological charge 
on a copy of it \cite{fp}. 
This is especially required at small values of $\beta$ where 
fluctuations at the scale of the correlation length
can be easily created as the lattice spacing is larger in 
physical units.  
In figure 1 we show the resulting
topological charge distribution on a set of 500  initial flat 
configurations updated during 36 heat-bath sweeps at $\beta=5.90$
and then frozen by 6 cooling steps. 
In the figure we clearly see a major peak at $Q=0$ and minor peaks
around non-zero integer values of $Q$, 
corresponding to configurations in which instantons or antiinstantons have 
been created.
The configurations representing these minor 
peaks must be discarded.
To do that we operate a cut  $|Q| < \delta$, at a $\delta$ of the order 
of the width of the major peak.  
The background in the figure gives an estimate of the systematic error.
We take as this systematic error the ratio 
$A_b/A_{Q=0}$ where $A_b$ is the
area of the background in the cut interval, 
and $A_{Q=0}$ is the area of the major peak.
This error is added to the statistical one by quadrature.

A similar procedure was developed for the determination of the
multiplicative renormalization $Z(\beta)$, as shown in fig. 2.

The simulations were performed on an APE QUADRICS machine and the 
Monte Carlo technique used was standard. 

For $a(\beta)$ we will use the formula
\begin{equation}
a(\beta) = {1 \over \Lambda_L} \left( {{8 \pi^2 \beta} \over
{33}} \right)^{51/121} \exp\left(-{{4 \pi^2 \beta} \over
33}\right).
\end{equation}
In this equation $\Lambda_L$ is an effective scale which in principle
depends on $\beta$ and becomes practically constant at large enough
$\beta$, when the two-loop approximation is good for asymptotic
scaling. As explained in next section, we will fix $\Lambda_L$ in
terms of $T_c$ and then remain in a small interval of $\beta$ such
that its dependence on $\beta$ can be neglected and eq. (2.4) is
valid. 

\section{The Monte Carlo results}

We measured the topological susceptibility at zero temperature on
a symmetric lattice $16^4$ and at finite temperature on a lattice
$32^3\times 8$. 

In figure 3 we plot the value of $(\chi)^{(1/4)}$ at zero
temperature versus $\beta$ for the three definitions 
$\chi_L^{(i)}$ ($i=0,1,2$) of the lattice
topological susceptibility. The corresponding data are listed
in Table I.
There is an excellent agreement between the three determinations. 
To fix the value of $\chi^{1/4}$ in physical units we can either use
\cite{boyd} $\beta_c(N_{\tau}=8)=6.0609(9)$ and eq. (2.4) and (1.11) which
gives $\Lambda_L$ in terms of $T_c$, 
or the determination of reference \cite{bali} $\Lambda_L=4.56(11)$.
The horizontal line is the
linear fit to the 2-smear data. It gives $(\chi)^{(1/4)} = 175(5)$
MeV,  and is consistent, within errors, with that of ref. 
\cite{digia3,np3,tepchi}. 
The error includes the uncertainty in $\Lambda_L$.

As a check of thermalization we show in figure 4 the
distribution of total topological charge for 5000 configurations at
$\beta=6.1$ for the 2-smeared charge. The figure was obtained by
applying 6 cooling steps on fully thermalized configurations.
The average topological charge is zero within errors, as it should be. 

In figure 5 the topological susceptibility 
$\chi \equiv (\chi^{(i)}_L - M^{(i)})/(Z^{(i)2} a^4)$
at the 
transition point is shown for the 1 and 2-smeared operator. $\chi$
drops by one 
order of magnitude from the confined to the deconfined phase. 
The results obtained with the two operators are compatible as they should.
The data for the 0-smeared operator have very large error
bars and are not shown in the figure.
The data have been plotted versus $T/T_c$ where $T_c$ is the
deconfining temperature. 
To determine $T/T_c$ we only need the ratio $a(\beta_c)/a(\beta)$ and
for that the two-loop expression is certainly a good approximation
within the small interval of $\beta$ used, where $\Lambda_L$ can be
considered as a constant.

The solid line in figure 5 corresponds to the value of $\chi$ at
zero-temperature and is consistent with the data below $T_c$
indicating that $\chi$ is practically $T$-independent in the confined
phase. 

In Table II we display the unrenormalized value
of the topological susceptibility $\chi_L^{(1)}$ and the
values for both $Z^{(1)}$ and $M^{(1)}$ 
as a function of $\beta$
for the measurements performed using the 1-smeared topological
charge density $Q_L^{(1)}$. The last column is the physical value for
the susceptibility $\chi/\Lambda_L^4$ obtained from equation 
(\ref{eq:defrenchii}) .
In Table III we show the analogous
set of data for the 2-smeared operator.

The quality of our 2-smeared operator can be appreciated by looking at
the numbers in the last column of Table IV; $M^{(2)}$, which includes
both the mixing to the action density and to the identity operator
(see eq. (2.4)), is $0.2\div 0.3$ times the subtracted signal
$\chi_L^{(2)} - M^{(2)}$. The deviations from constant of that ratio
allows to estimate $P^{(2)}(\beta)$. At $\beta=6.1$, 
$P^{(2)}(\beta)/\chi_L^{(2)}
 \lesssim 3\%$. Going to lower $\beta$'s this ratio decreases very
rapidly; at larger $\beta$'s both the terms proportional to $a^4$ in 
eq. (2.3) die off exponentially and $P^{(2)}(\beta)$ becomes dominant. 
A similar analysis on the data of Table V for the 1-smeared operator
shows a much lower quality. For the non-improved operator the 
ratio $M^{(0)}/(\chi_L^{(0)} - M^{(0)})$ is
much larger than 1, and $M^{(0)}(\beta)$ is dominated by $P^{(0)}(\beta)$.

A preliminary estimate of the mixing to the gluon condensate $G_2$
across $T_c$ shows that $G_2$ is much smoother than $\chi$ at
$T_c$ and compatible with a constant. Work is in progress on this
point to reduce the error bars.

\section{Concluding remarks}

We have used an improved operator for the topological charge density 
to determine the topological susceptibility of $SU(3)$ pure gauge theory 
at zero temperature and its behaviour through $T_c$. 

Lattice artefacts are strongly suppressed with respect to the ordinary 
definition, eq. (1.11), 
and are anyhow removed by non-perturbative methods. 
Moreover statistical fluctuations are drastically reduced.

Our main results are: 

 1) At $T = 0$ the Witten-Veneziano solution \cite{wit2,vene2}
of the $U_A(1)$ problem is 
confirmed. The value of $\chi$ is more precise than in previous 
determinations and agrees with them within errors.
  
 2) $\chi$ drops to zero at $T_c$.

The method used rests on basic concepts in field theory, and on the assumption
 that lattice is a legitimate regularization of it. No 
additional assumptions are needed.

\section{Acknowledgements}

We thank Graham Boyd and Enrico Meggiolaro
for useful discussions. A.D.G. acknowledges an interesting discussion
with Heinrich Leutwyler.
B.A. acknowledges financial support from an 
I.N.F.N. contract.


\newpage

\noindent{\bf Figure captions}

\begin{enumerate}

\item[Figure 1.] Distribution of topological charge $Q$ for a set of
500 configurations obtained by 36 heat-bath updatings of the flat
configuration and 6 cooling sweeps. $\beta=5.90$.

\item[Figure 2.] Distribution of topological charge $Q$ for a set of
2000 configurations obtained by 15 heat-bath updatings of a
1-instanton 
configuration and 6 cooling sweeps. $\beta=5.75$.

\item[Figure 3.] $\chi$ at $T=0$.
The 
straight line is the result of the linear fit of the 2-smeared data.
The improvement from $Q^{(0)}_L$ to $Q^{(2)}_L$ is clearly visible.

\item[Figure 4.] Topological charge distribution for an ensemble of
5000
fully thermalized configurations at $\beta=6.1$, (2-smeared
operator).

\item[Figure 5.] $\chi/\Lambda_L^4$ versus $T/T_c$ across
the deconfining phase transition.
The horizontal band is the determination at $T=0$ of figure 2.

\end{enumerate}

\vskip 2cm

\noindent{\bf Table captions}

\vskip 5mm

\begin{enumerate}

\item[Table I.] $\chi^{1/4}/\Lambda_L$ 
from the 0,1 and 2-smeared operators on a $16^4$ lattice.

\item[Table II.] $T/T_c$, $\chi_L^{(1)}$, $M^{(1)}$ and $Z^{(1)}$ as a
function of $\beta$ for the 1-smeared operator.

\item[Table III.] $T/T_c$, $\chi_L^{(2)}$, $M^{(2)}$ and $Z^{(2)}$ as a
function of $\beta$ for the 2-smeared operator.

\item[Table IV.] Data for $\chi_L^{(2)}$ and $M^{(2)}$ for the
2-smeared operator.

\item[Table V.] Data for $\chi_L^{(1)}$ and $M^{(1)}$ for the
1-smeared operator.

\end{enumerate}

\newpage

\vskip 2cm

\centerline{\bf Table I}

\vskip 5mm

\moveright 1.0 in
\vbox{\offinterlineskip
\halign{\strut
\vrule \hfil\quad $#$ \hfil \quad & 
\vrule \hfil\quad $#$ \hfil \quad & 
\vrule \hfil\quad $#$ \hfil \quad & 
\vrule \hfil\quad $#$ \hfil \quad \vrule \cr
\noalign{\hrule}
\beta & 
\chi^{(1/4)}_{\rm 0-smear}/\Lambda_L &
\chi^{(1/4)}_{\rm 1-smear}/\Lambda_L &
\chi^{(1/4)}_{\rm 2-smear}/\Lambda_L \cr
\noalign{\hrule}
5.90 & 36.7(12.9) & 39.1(1.1) & 38.8(0.8) \cr
\noalign{\hrule}
6.00 & 40.1(14.8) & 39.1(1.1) & 38.4(0.8) \cr
\noalign{\hrule}
6.10 & 43.1(14.4) & 39.0(1.2) & 38.1(0.8) \cr
\noalign{\hrule}
}}

\vskip 2cm

\centerline{\bf Table II}

\vskip 5mm

\moveright 0.4 in
\vbox{\offinterlineskip
\halign{\strut
\vrule \hfil\quad $#$ \hfil \quad & 
\vrule \hfil\quad $#$ \hfil \quad & 
\vrule \hfil\quad $#$ \hfil \quad & 
\vrule \hfil\quad $#$ \hfil \quad & 
\vrule \hfil\quad $#$ \hfil \quad & 
\vrule \hfil\quad $#$ \hfil \quad \vrule \cr
\noalign{\hrule}
\beta & 
T/T_c &
10^5 \times \chi_L^{(1)} &
10^5 \times M^{(1)} &
Z^{(1)} &
10^6 \times \chi/\Lambda_L^4 \cr
\noalign{\hrule}
5.90 & 0.834 & 2.35(7) & 0.88(6) & 0.36(2) & 2.46(32) \cr
\noalign{\hrule}
6.00 & 0.934 & 1.55(5) & 0.62(3) & 0.39(2) & 2.07(25) \cr
\noalign{\hrule}
6.06 & 0.999 & 1.05(3) & 0.53(4) & 0.40(2) & 1.38(19) \cr
\noalign{\hrule}
6.10 & 1.045 & 0.728(21) & 0.48(3) & 0.42(2) & 0.76(13) \cr
\noalign{\hrule}
6.22 & 1.197 & 0.428(12) & 0.34(2) & 0.45(2) & 0.38(11) \cr
\noalign{\hrule}
6.30 & 1.310 & 0.334(9) & 0.314(15) & 0.48(2) & 0.11(10) \cr
\noalign{\hrule}
6.36 & 1.402 & 0.293(10) & 0.281(10) & 0.49(2) & 0.08(10) \cr
\noalign{\hrule}
}}

\newpage

\centerline{\bf Table III}

\vskip 5mm

\moveright 0.4 in
\vbox{\offinterlineskip
\halign{\strut
\vrule \hfil\quad $#$ \hfil \quad & 
\vrule \hfil\quad $#$ \hfil \quad & 
\vrule \hfil\quad $#$ \hfil \quad & 
\vrule \hfil\quad $#$ \hfil \quad & 
\vrule \hfil\quad $#$ \hfil \quad & 
\vrule \hfil\quad $#$ \hfil \quad \vrule \cr
\noalign{\hrule}
\beta & 
T/T_c &
10^5 \times \chi_L^{(2)} &
10^5 \times M^{(2)} &
Z^{(2)} &
10^6 \times \chi/\Lambda_L^4 \cr
\noalign{\hrule}
5.90 & 0.834 & 3.27(9) & 0.71(6) & 0.49(2) & 2.07(18) \cr
\noalign{\hrule}
6.00 & 0.934 & 2.04(7) & 0.46(3) & 0.51(2) & 2.00(18) \cr
\noalign{\hrule}
6.06 & 0.999 & 1.31(4) & 0.38(3) & 0.53(2) & 1.54(15) \cr
\noalign{\hrule}
6.10 & 1.045 & 0.76(2) & 0.33(2) & 0.55(2) & 0.85(9) \cr
\noalign{\hrule}
6.22 & 1.197 & 0.320(9) & 0.227(14) & 0.58(2) & 0.32(6) \cr
\noalign{\hrule}
6.30 & 1.310 & 0.227(6) & 0.197(9) & 0.60(2) & 0.10(4) \cr
\noalign{\hrule}
6.36 & 1.402 & 0.184(6) & 0.168(7) & 0.622(13) & 0.07(4) \cr
\noalign{\hrule}
}}

\vskip 2cm

\centerline{\bf Table IV}

\vskip 5mm

\moveright 1.0 in
\vbox{\offinterlineskip
\halign{\strut
\vrule \hfil\quad $#$ \hfil \quad & 
\vrule \hfil\quad $#$ \hfil \quad & 
\vrule \hfil\quad $#$ \hfil \quad & 
\vrule \hfil\quad $#$ \hfil \quad \vrule \cr
\noalign{\hrule}
\beta & 
10^5 \times \chi_L^{(2)} &
10^5 \times M^{(2)} &
M^{(2)}/(\chi_L^{(2)} - M^{(2)}) \cr
\noalign{\hrule}
5.90 & 3.51(7) & 0.71(6) & 0.253(23) \cr
\noalign{\hrule}
6.00 & 2.18(5) & 0.46(3) & 0.270(20) \cr
\noalign{\hrule}
6.10 & 1.39(3) & 0.33(2) & 0.315(22) \cr
\noalign{\hrule}
}}

\vskip 2cm

\centerline{\bf Table V}

\vskip 5mm

\moveright 1.0 in
\vbox{\offinterlineskip
\halign{\strut
\vrule \hfil\quad $#$ \hfil \quad & 
\vrule \hfil\quad $#$ \hfil \quad & 
\vrule \hfil\quad $#$ \hfil \quad & 
\vrule \hfil\quad $#$ \hfil \quad \vrule \cr
\noalign{\hrule}
\beta & 
10^5 \times \chi_L^{(1)} &
10^5 \times M^{(1)} &
M^{(1)}/(\chi_L^{(1)} - M^{(1)}) \cr
\noalign{\hrule}
5.90 & 2.48(5) & 0.87(4) & 0.540(32) \cr
\noalign{\hrule}
6.00 & 1.64(4) & 0.60(2) & 0.576(29) \cr
\noalign{\hrule}
6.10 & 1.12(2) & 0.474(13) & 0.734(34) \cr
\noalign{\hrule}
}}

\begin{figure}
{\centerline{\bf Figure 1}}
\vspace{-1.5cm}
\hspace{-1cm}
\epsffile{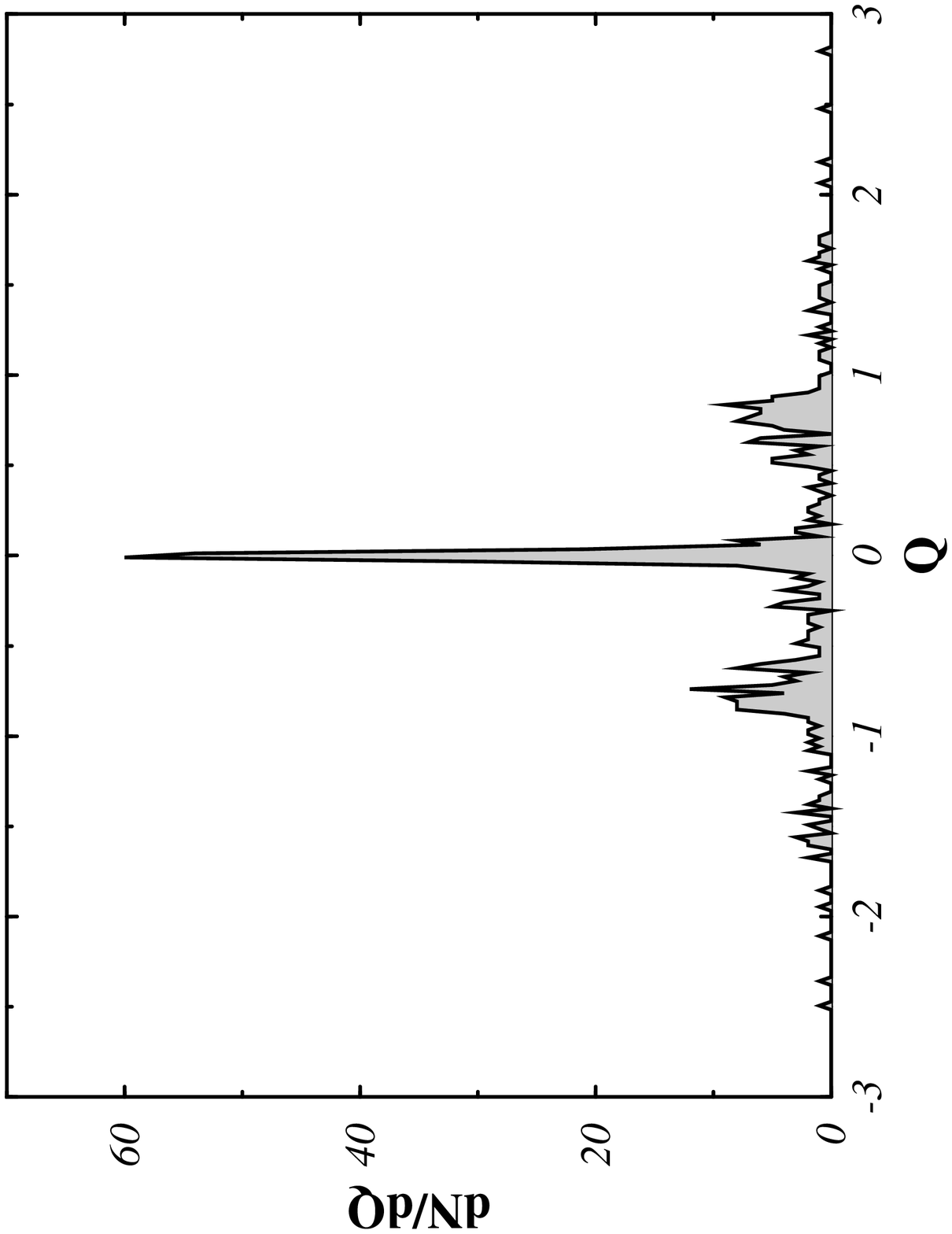}
\end{figure}

\newpage



\begin{figure}
{\centerline{\bf Figure 2}}
\vspace{-1.5cm}
\hspace{-1cm}
\epsffile{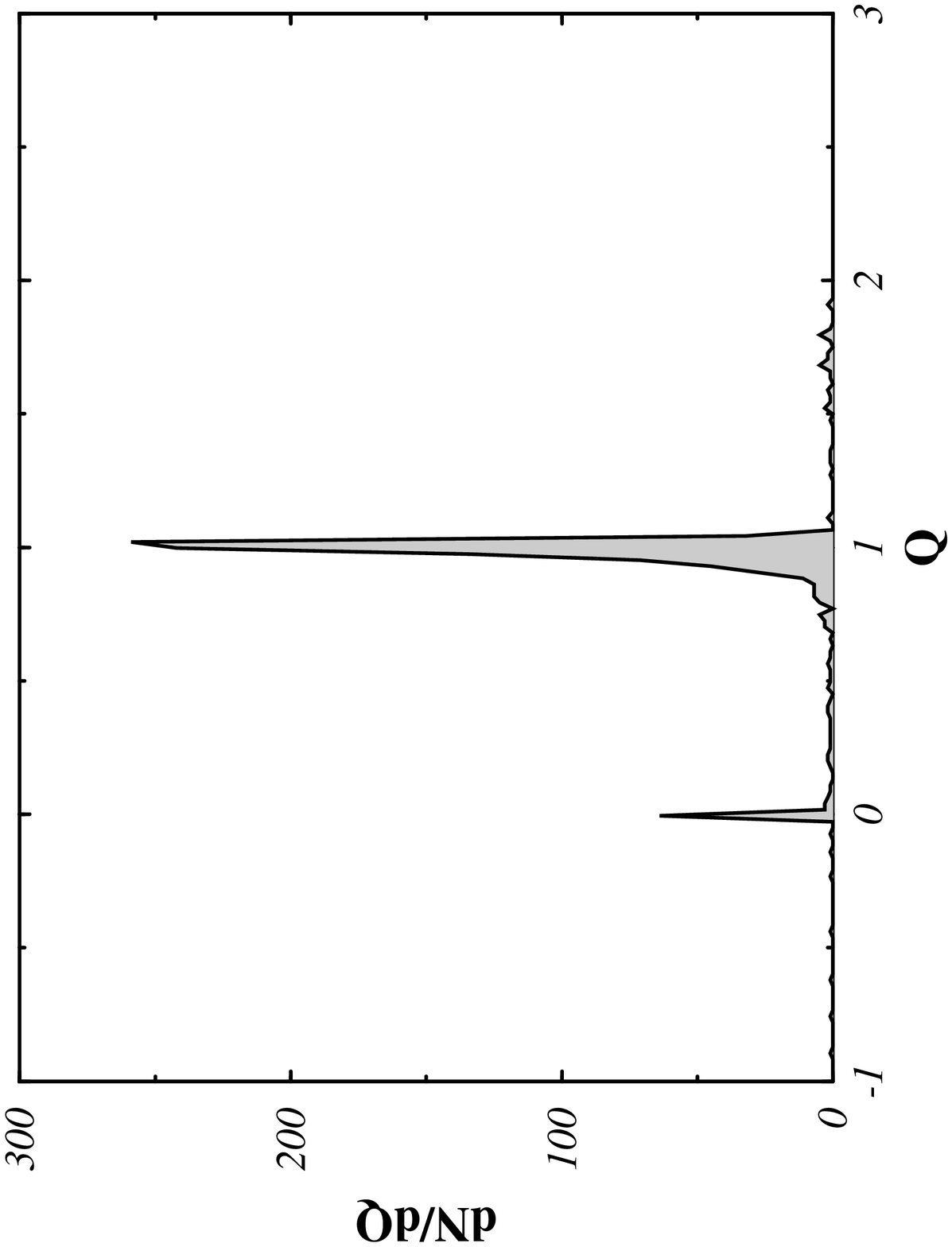}
\end{figure}

\newpage

\begin{figure}
{\centerline{\bf Figure 3}}
\vspace{-0.4cm}
\hspace{-1cm}
\epsffile{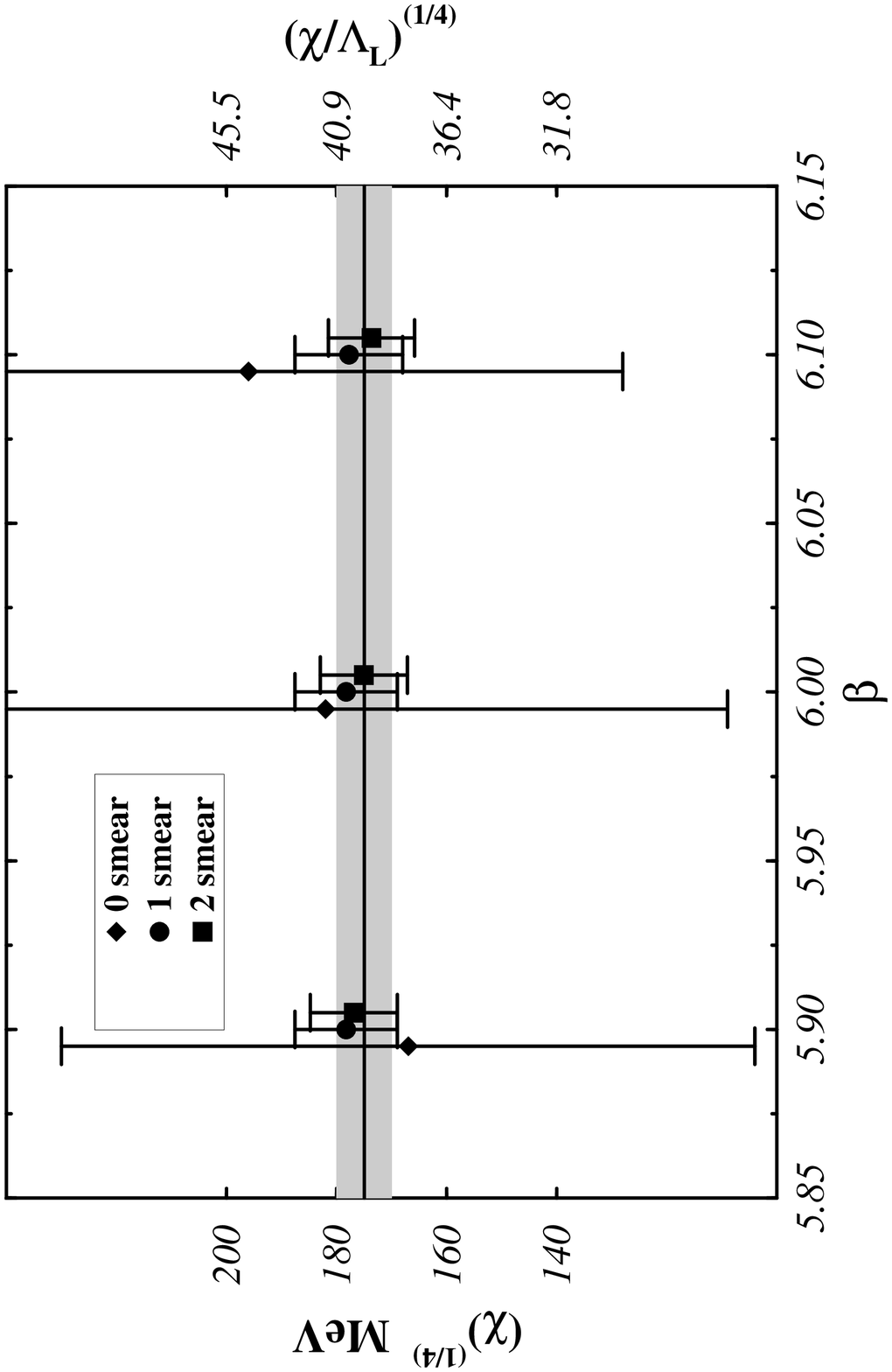}
\end{figure}

\newpage

\begin{figure}
{\centerline{\bf Figure 4}}
\vspace{-1.5cm}
\hspace{-1cm}
\epsffile{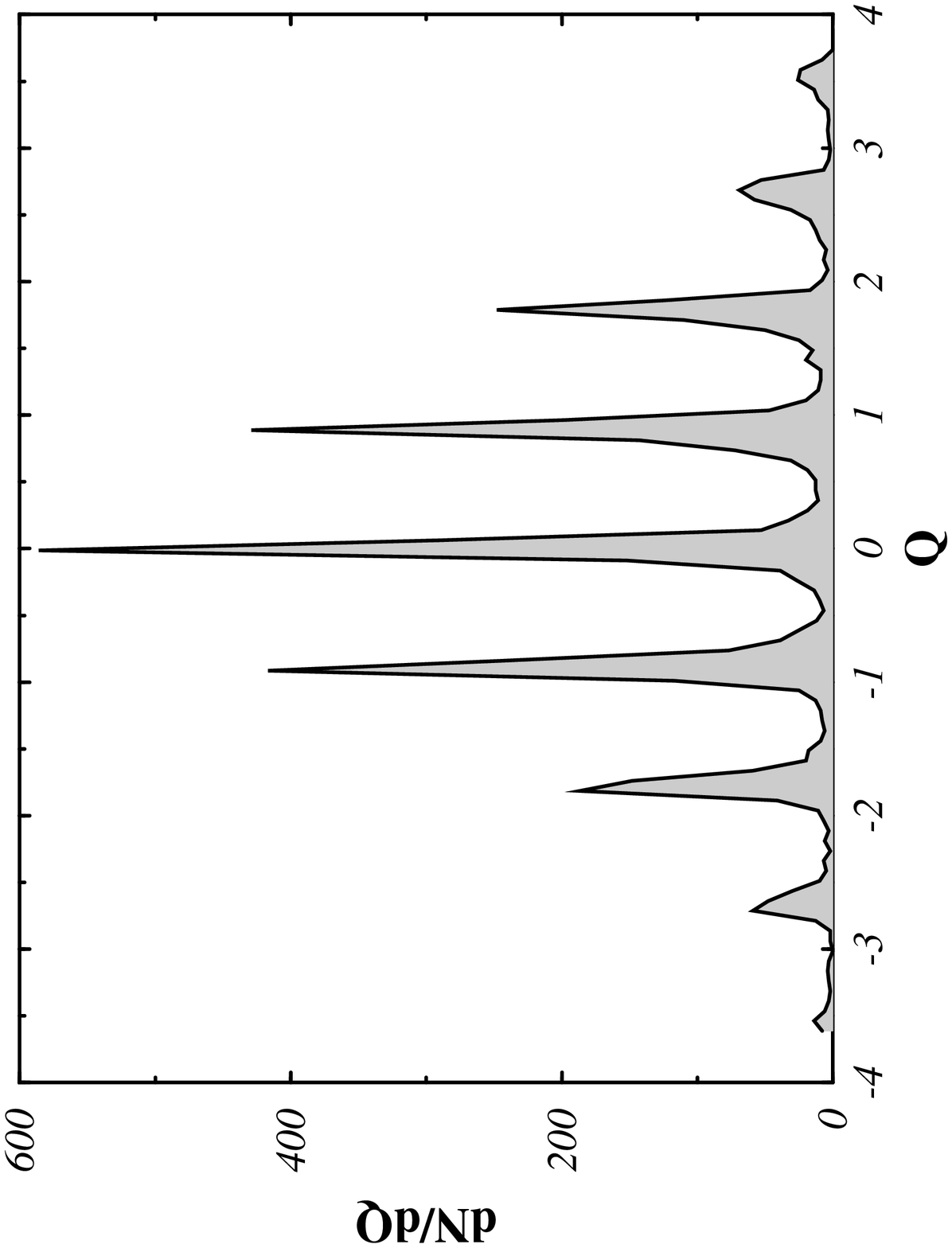}
\end{figure}

\newpage

\begin{figure}
{\centerline{\bf Figure 5}}
\vspace{-0.4cm}
\hspace{-1cm}
\epsffile{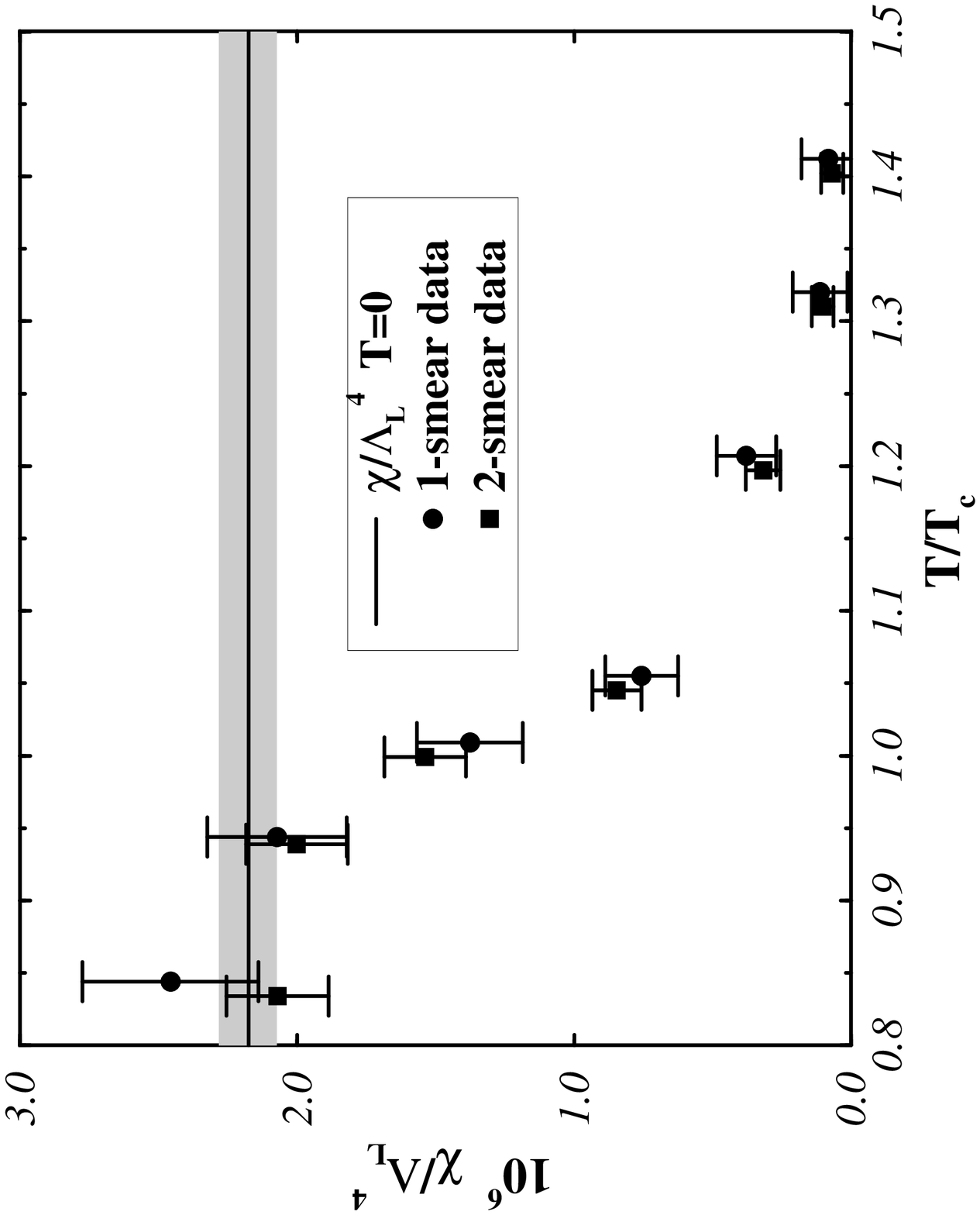}
\end{figure}

\newpage

\vskip 2cm

\centerline{{\bf {\underline{APPENDIX}}}}
\centerline{{\bf Erratum to:} }
\centerline{{\it ``Topological susceptibility at zero and finite T  }}
\centerline{{\it in SU(3) Yang--Mills theory''}}
\vskip 3mm
\centerline{\bf [Nuclear Physics B494 (1997) pg. 281]}
\vskip 3mm
\centerline{B. All\'es, M. D'Elia, A. Di Giacomo}

\vskip 7mm

\noindent In the computation of Table 1 and of the last columns of Tables~2
and~3 some trivial numerical error was made. These numbers are computed
from the lattice data of Tables~2--5 by use of Eq.(1.10) and~(2.5). The
correct tables are displayed below showing a marginal violation of
asymptotic scaling.

It is therefore better to present our results in terms of the scaling
ratio $\chi^{1/4}/T_{\rm c}$, discarding the approximation Eq.~(2.5) and
using instead the numerical determinations of $T_{\rm c}$ at 
different $\beta$ of Ref.~\cite{karsch}. The results are shown in 
Figure~3 and~5 below.
The conclusions of the paper stay unaltered.

The MeV units in Fig.~3 have been
obtained by using the ratio $T_{\rm c}/\sqrt{\sigma}=0.629(3)$ 
from Ref.~\cite{karsch} and the value $\sqrt{\sigma}=420$~MeV. 
From the 2-smeared data we derive $\chi^{1/4}=170(7)$~MeV. 
If the ratio $T_{\rm c}/\sqrt{\sigma}$
is taken from~\cite{teper}, this result becomes $\chi^{1/4}=174(7)$~MeV.
The new Figure~5 displays the scaling ratio $\chi/T_{\rm c}^4$.
The temperature axis in this Figure has been
determined again by using $T_{\rm c}$ from Ref.~\cite{karsch}.

\vskip 5mm

\centerline{\bf Table 1}

\vskip 1mm

\moveright 0.5 in
\vbox{\offinterlineskip
\halign{\strut
\vrule \hfil\quad $#$ \hfil \quad & 
\vrule \hfil\quad $#$ \hfil \quad & 
\vrule \hfil\quad $#$ \hfil \quad & 
\vrule \hfil\quad $#$ \hfil \quad \vrule \cr
\noalign{\hrule}
\beta &
\chi^{(1/4)}_{\rm 0-smear}/\Lambda_L &
\chi^{(1/4)}_{\rm 1-smear}/\Lambda_L &
\chi^{(1/4)}_{\rm 2-smear}/\Lambda_L \cr
\noalign{\hrule}
5.90 & 42.3(10.4) & 40.2(1.2)  &  39.6(0.9) \cr
\noalign{\hrule}
6.00 & 37.8(8.2)  & 38.8(1.1)  &  38.4(0.8) \cr
\noalign{\hrule}
6.10 & 40.6(7.8)  & 37.1(1.0)  &  36.7(0.7) \cr
\noalign{\hrule}
}}

\vskip 1cm


\newpage
\centerline{\bf {\rm First and last columns in} Table 2}

\vskip 1mm

\moveright 1.8 in
\vbox{\offinterlineskip
\halign{\strut
\vrule \hfil\quad $#$ \hfil \quad & 
\vrule \hfil\quad $#$ \hfil \quad \vrule \cr
\noalign{\hrule}
\beta &
10^{-6}\times \chi/\Lambda_L^4 \cr
\noalign{\hrule}
5.90 & 2.39(30) \cr
\noalign{\hrule}
6.00 & 2.02(24) \cr
\noalign{\hrule}
6.06 & 1.41(20) \cr
\noalign{\hrule}
6.10 & 0.73(13) \cr
\noalign{\hrule}
6.22 & 0.39(11) \cr
\noalign{\hrule}
6.30 & 0.11(10) \cr
\noalign{\hrule}
6.36 & 0.08(10) \cr
\noalign{\hrule}
}}

\vskip 1cm

\centerline{\bf {\rm First and last columns in} Table 3}

\vskip 1mm

\moveright 1.8 in
\vbox{\offinterlineskip
\halign{\strut
\vrule \hfil\quad $#$ \hfil \quad & 
\vrule \hfil\quad $#$ \hfil \quad \vrule \cr
\noalign{\hrule}
\beta &
10^{-6}\times \chi/\Lambda_L^4 \cr
\noalign{\hrule}
5.90 & 2.24(21) \cr
\noalign{\hrule}
6.00 & 2.00(18) \cr
\noalign{\hrule}
6.06 & 1.43(13) \cr
\noalign{\hrule}
6.10 & 0.74(7) \cr
\noalign{\hrule}
6.22 & 0.25(5) \cr
\noalign{\hrule}
6.30 & 0.11(4) \cr
\noalign{\hrule}
6.36 & 0.07(4) \cr
\noalign{\hrule}
}}

\vskip 5mm

\newpage

\begin{figure}
{\centerline{\bf Figure 3}}
\vspace{-3.5cm}
\centerline{\ewxy{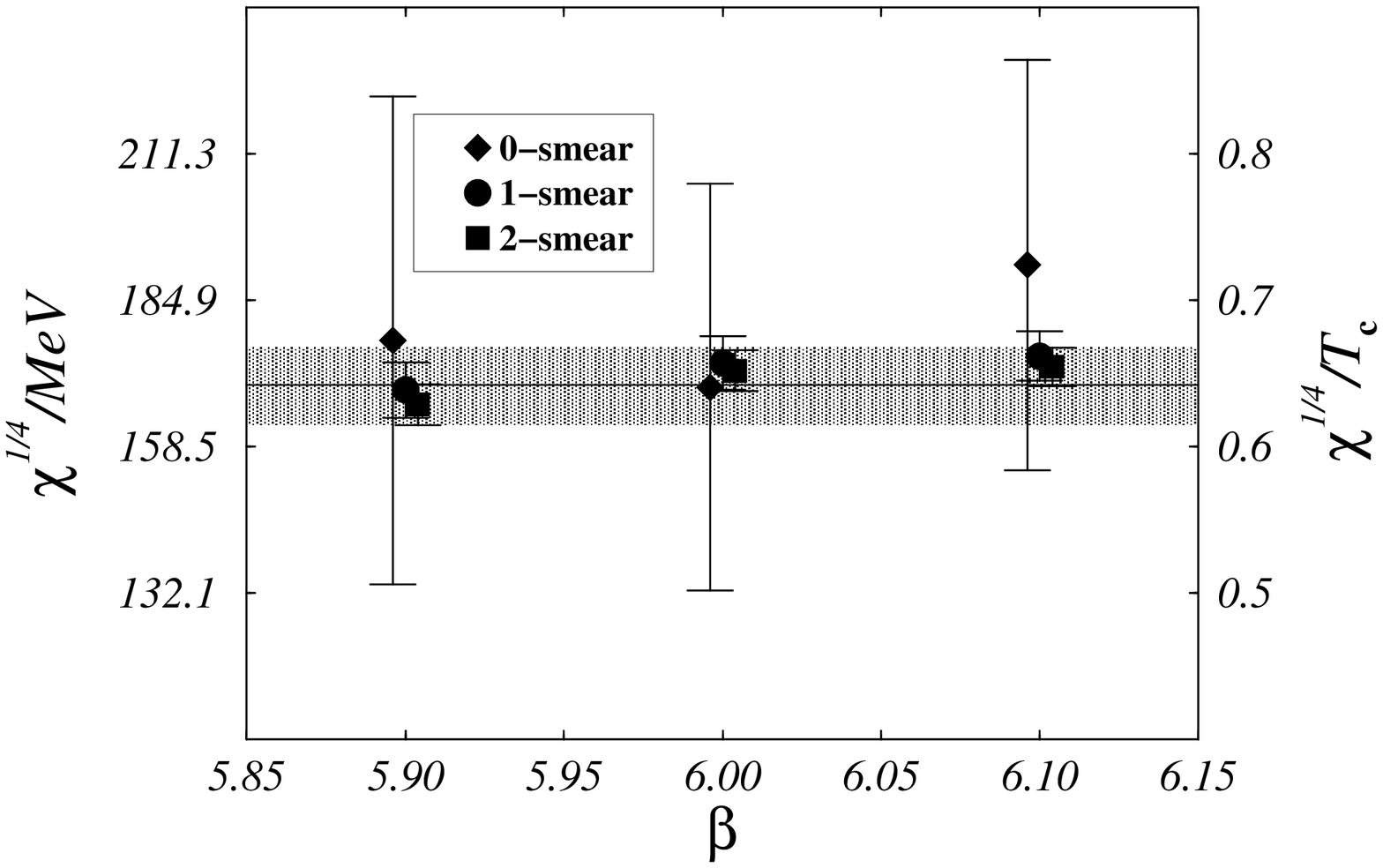}{400pt}}
\protect\label{fig:glprop}
\end{figure}

\vskip 5mm

\begin{figure}
{\centerline{\bf Figure 5}}
\vspace{-3.cm}
\centerline{\ewxy{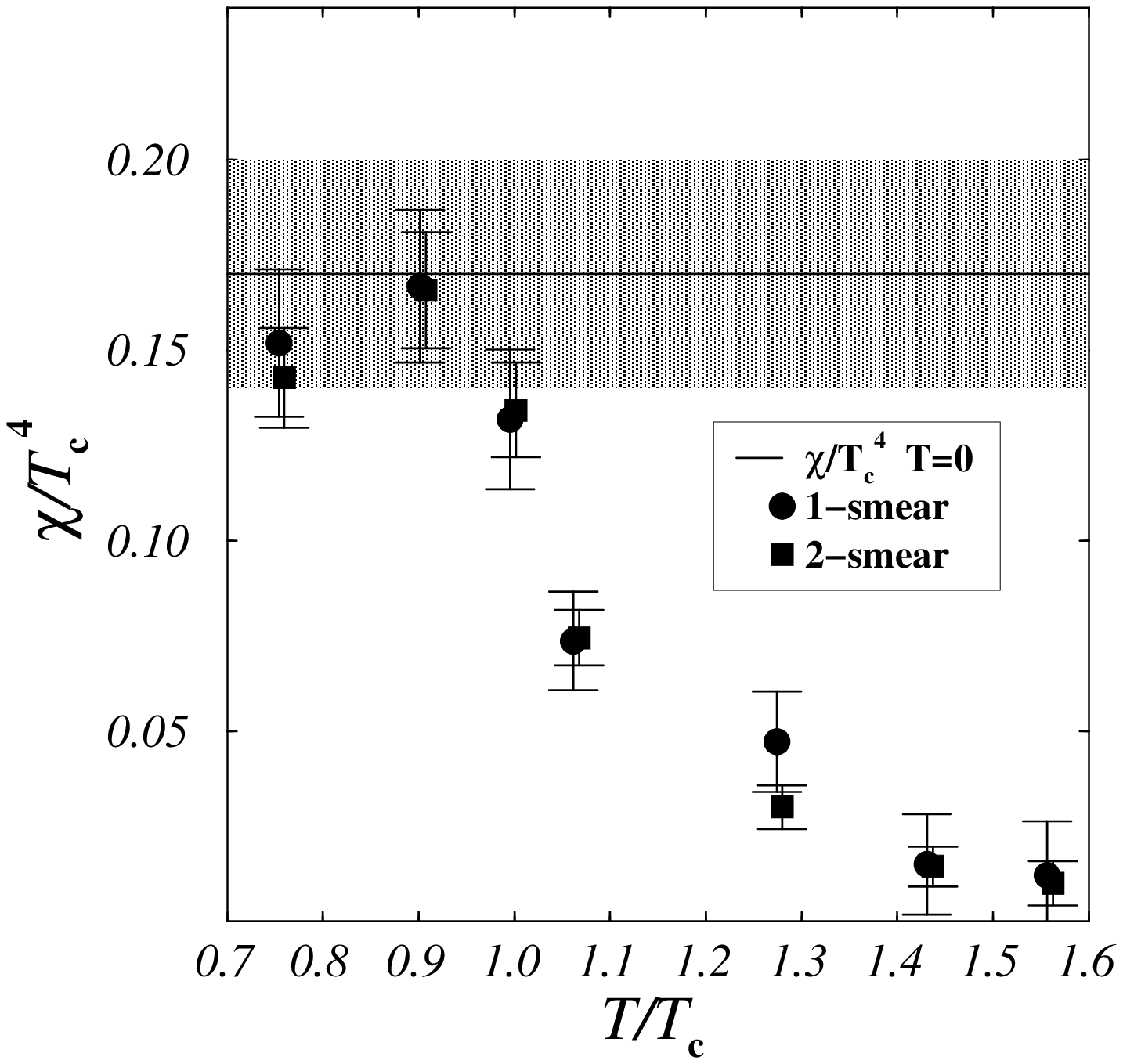}{400pt}}
\protect\label{fig:glpropp}
\end{figure}

\noindent We are grateful to Ettore Vicari for pointing out to us the
inexact numbers in Table~1.



\begin{thebibliography}{99}
\bibitem{hoft1} G. t'Hooft, Phys. Rev. Lett. {\bf 37} (1976) 8.
\bibitem{wein} S. Weinberg, Phys. Rev. {\bf D11} (1975) 3583.
\bibitem{wit1} E. Witten, Nucl. Phys. {\bf B149} (1979) 285.
\bibitem{hoft2} G. t'Hooft, Nucl. Phys. {\bf B72} (1974) 461.
\bibitem{vene1} G. Veneziano, Nucl. Phys. {\bf B117} (1976) 519.
\bibitem{wit2} E. Witten, Nucl. Phys. {\bf B156} (1979) 269.
\bibitem{vene2} G. Veneziano, Nucl. Phys. {\bf B159} (1979) 213.
\bibitem{digia1} For a review see e.g. A. Di Giacomo, Nucl. Phys. 
(Proc. Suppl.)  {\bf B23} (1981) 191.
\bibitem{fuku} M. Fukugita, Y. Kuramashi, M. Okawa, A. Ukawa, Phys. Rev. 
{\bf D51} (1995) 3952.
\bibitem{boyd} G. Boyd, J. Engels, F. Karsch, E. Laermann,
C. Legeland, M. L\"utgemeier and B. Petersson, Bielefeld preprint
BI-TP 96/04 and hep-lat 9602007.
\bibitem{shu1} See e.g. E. Shuryak, Comments in Nuclear Particle 
Physics {\bf 21} (1994) 235.
\bibitem{pisar} R. D. Pisarski, L.G.Yaffe, Phys. Lett. {\bf B97} (1980) 110.
\bibitem{tep2} M. Teper, Phys. Lett. {\bf B171} (1986) 81, 86.
\bibitem{digia2} A. Di Giacomo, E. Meggiolaro, H. Panagopoulos, Phys. Lett. 
{\bf B277} (1992) 491.
\bibitem{z1} M. Campostrini, A. Di Giacomo and H. Panagopoulos,
Phys. Lett. {\bf B212} (1988) 206.
\bibitem{tutti} M. Campostrini, A. Di Giacomo, H. Panagopoulos and
E. Vicari, Nucl. Phys. {\bf B329} (1990) 683.
\bibitem{crew} R. J. Crewther, Riv. Nuovo Cim. {\bf 2} (1979) 63.
\bibitem{divec} P. Di Vecchia, K. Fabricius, G.C. Rossi, G. Veneziano, 
Nucl. Phys. {\bf B192} (1981) 392.
\bibitem{digia3} M. Campostrini, A. Di Giacomo, Y. G\"und\"u\c c, 
M. P. Lombardo, H. Panagopoulos, R. Tripiccione, Phys. Lett. {\bf B252} 
(1990) 436.  
\bibitem{np1} A. Di Giacomo and E. Vicari, Phys. Lett. {\bf B275}
(1992) 429.
\bibitem{np3bis} B. All\'es, M. Campostrini, A. Di Giacomo,
Y. G\"und\"u\c c and E. Vicari, Phys. Rev. {\bf D48} (1993) 2284.
\bibitem{np3} B. All\'es, M. Campostrini, A. Di Giacomo,
Y. G\"und\"u\c c and E. Vicari, Nucl. Phys. (Proc. Suppl.) {\bf B34}
(1994) 504.
\bibitem{lus} M. L\"uscher, Commun. Math. Phys. {\bf 85} (1982) 28.
\bibitem{kron} A. S. Kronfeld, A. Laursen, G. Schierholz, U. Wiese, 
Nucl. Phys. {\bf B292} (1987) 330.
\bibitem{tep3} Pugh, M. Teper, Phys. Lett. {\bf B218} (1989) 326.
\bibitem{hperfect} T. DeGrand, A. Hasenfratz, P. Hasenfratz and
F. Niedermayer, Nucl. Phys. {\bf B454} (1995) 587.
\bibitem{christou} C. Christou, A. Di Giacomo, H. Panagopoulos and
E. Vicari, Phys. Rev. {\bf D53} (1996) 2619.
\bibitem{fp} F. Farchioni and A. Papa, Nucl. Phys. {\bf B431} (1994)
686. 
\bibitem{tepchi} M. Teper, Phys. Lett. {\bf B202} (1988) 553.
\bibitem{bali} G. S. Bali and K. Schilling, Phys. Rev. {\bf D47}
(1993) 661.
\end{thebibliography}

\begin{thebibliography}{99}
\bibitem{karsch} G. Boyd, J. Engels, F.~Karsch, E.~Laermann, C.~Legeland, 
M.~L\"utgemeier, B.~Petersson, Nucl. Phys. B469 (1996) 419.
\bibitem{teper} B.~Lucini, M.~Teper, U.~Wenger, hep-lat/0307017.
\end{thebibliography}
\end{document}